\documentclass[conference]{IEEEtran}
\IEEEoverridecommandlockouts
\usepackage{amsmath,amssymb,amsfonts}
\usepackage{algorithmic}
\usepackage{graphicx}
\usepackage{textcomp}
\usepackage{xcolor}
\def\BibTeX{{\rm B\kern-.05em{\sc i\kern-.025em b}\kern-.08em
    T\kern-.1667em\lower.7ex\hbox{E}\kern-.125emX}}

\usepackage{xurl}
\usepackage[hidelinks]{hyperref}
\usepackage{booktabs}
\usepackage{multirow}
\usepackage{array}
\usepackage{soul}

\usepackage{amssymb}            %
\usepackage{pifont}             %

\newcommand{\cmark}{\checkmark}       %
\newcommand{\xmark}{\ding{55}}        %

\definecolor{mynavy}{RGB}{128, 200, 255}
\sethlcolor{white}
\soulregister\cite7
\soulregister\ref7
\soulregister\pageref7

\newcommand{\hbl}[1]{\hl{#1}} %

\usepackage[maxnames=1, minnames=1, backend=biber, giveninits=true, sorting=none]{biblatex}
\DeclareSourcemap{
    \maps[datatype=bibtex]{
        \map{
            \step[fieldset=ISSN, null]
            \step[fieldset=ISBN, null]
            \step[fieldset=urldate, null]
            \step[fieldset=note, null]
            \step[fieldset=eventtitle, null]
            \step[fieldset=publisher, null]
            \step[fieldset=pages, null]
            \step[fieldset=doi, null]
            \step[fieldset=language, null]
        }
    }
}

\newbibmacro*{conditionally-print-doi}{%
  \ifboolexpr{
    test {\iffieldundef{journal}} 
    and 
    test {\iffieldundef{booktitle}}
    and
    test {\iffieldundef{journaltitle}}
  }
  {\printfield{doi}}{}
}

\renewbibmacro*{doi+eprint+url}{
    \usebibmacro{conditionally-print-doi}
}

\begin{document}

\title{CloudFormer: An Attention-based
Performance Prediction for Public Clouds with Unknown Workload
}
\author{\IEEEauthorblockN{Amirhossein Shahbazinia}
\IEEEauthorblockA{\textit{ESL, EPFL}\\
Lausanne, Switzerland \\
amirhossein.shahbazinia@epfl.ch}
\and
\IEEEauthorblockN{Darong Huang}
\IEEEauthorblockA{\textit{ESL, EPFL}\\
Lausanne, Switzerland \\
darong.huang@epfl.ch}
\and
\IEEEauthorblockN{Luis Costero}
\IEEEauthorblockA{\textit{UCM}\\
Madrid, Spain \\
lcostero@ucm.es}
\and
\and
\IEEEauthorblockN{David Atienza}
\IEEEauthorblockA{\textit{ESL, EPFL}\\
Lausanne, Switzerland \\
david.atienza@epfl.ch}
\and
}

\maketitle

\begin{abstract}
Cloud platforms are increasingly relied upon to host diverse, resource-intensive workloads due to their scalability, flexibility, and cost-efficiency. In multi-tenant cloud environments, virtual machines are consolidated on shared physical servers to improve resource utilization. While virtualization guarantees resource partitioning for CPU, memory, and storage, it cannot ensure performance isolation. Competition for shared resources such as last-level cache, memory bandwidth, and network interfaces often leads to severe performance degradation. Existing management techniques, including VM scheduling and resource provisioning, require accurate performance prediction to mitigate interference. However, this remains challenging in public clouds due to the black-box nature of VMs and the highly dynamic nature of workloads.
To address these limitations, we propose CloudFormer, a dual-branch Transformer-based model designed to predict VM performance degradation in black-box environments. CloudFormer jointly models temporal dynamics and system-level interactions, using 206 system metrics at a one-second resolution across both static and dynamic scenarios. This design enables the model to capture transient interference effects and adapt to varying workload conditions without scenario-specific tuning. Complementing the methodology, we provide a fine-grained dataset that significantly expands the temporal resolution and metric diversity compared to existing benchmarks.
Experimental results demonstrate that CloudFormer consistently outperforms state-of-the-art baselines across multiple evaluation metrics, achieving robust generalization across diverse and previously unseen workloads. In particular, CloudFormer achieves a mean absolute error (MAE) of just 7.8\%, representing a substantial improvement in predictive accuracy and outperforming existing methods at least by 28\%.
\end{abstract}


\section{Introduction}

Global end-user spending on public cloud services has increased, exceeding 700 billion dollars in 2025, with sustained double digit growth expected in the coming years~\cite{gartner_gartner_2024}. This expanding demand brings significant energy implications: global data center electricity consumption reached an estimated 240-340 terawatt hours (TWh) in 2022, or roughly 1.0–1.3\% of global electricity demand, and US consumption hit 176 TWh (4.4\%) in 2023, rising rapidly thereafter~\cite{mytton_data_2022}. The International Energy Agency projects that the global demand for power from data centers will more than double to approximately 945 TWh by 2030~\cite{iea_ai_2025}, intensifying the need for more efficient resource allocation strategies and robust mechanisms to mitigate performance interference.

Virtualization technologies such as Intel VT and \mbox{AMD-V}~\cite{neiger_intel_2006, amd_amd_nodate} enable providers to consolidate multiple virtual machines (VMs) on a single physical server, improving hardware utilization~\cite{cortez_resource_2017}. Although resource isolation between VMs is guaranteed for dedicated CPU cores, memory allocations, and disk partitions, performance isolation remains a persistent challenge. VMs still compete for shared resources, such as last-level cache (LLC), memory bandwidth, and network interfaces, among others. This contention can significantly degrade VM performance, particularly under multi-tenant workloads~\cite{kim_virtual_2013}.  

In practice, cloud providers operate under strict privacy constraints: VMs are treated as black-boxes, with no access to the application source code or internal runtime metrics. Performance monitoring is therefore limited to host-level hardware counters, which complicates runtime performance prediction~\cite{wood_sandpiper_2009}. The challenge intensifies under dynamic workloads, where performance variation may arise from both interference due to co-located tenants and intrinsic workload variation. 
Disentangling these effects is not trivial, yet essential for guiding resource management decisions. Accurate performance forecasting serves as a critical primitive for intelligent orchestration, enabling schedulers to optimize VM placement to minimize interference~\cite{delimitrou_paragon_2013, cortez_resource_2017}, trigger resource throttling mechanisms for noisy neighbors~\cite{gan_seer_2019}, and ensure strict Service Level Agreement (SLA) compliance through proactive migration~\cite{novakovic_deepdive_2013}. Without a reliable quantitative signal, these downstream systems are forced to react only after QoS violations have already occurred.

Despite significant research on VM performance monitoring and prediction~\cite{kim_virtual_2013, anwar_game-theoretic_2021, akbar_game-based_2021}, existing methods face two key limitations. First, from a methodological point of view, prior models often rely on scenario-specific configurations (e.g., CPU- or network-intensive workloads), limiting their generalization to diverse and evolving cloud environments~\cite{pons_cloud_2023, buchaca_sequence--sequence_2020, masouros_rusty_2021, gan_seer_2019}. These methods primarily focus on static workload scenarios as they are easier to control and reproduce, avoiding the complexities introduced by dynamic workload patterns and their temporal variability. Second, from a data perspective, publicly available datasets either lack fine-grained temporal resolution or provide only limited metrics or data~\cite{wilkes_more_2011, tian_characterizing_2019, cortez_resource_2017, huang_cloudprophet_2024}, restricting their utility for training robust predictive models.

To address these limitations, we propose CloudFormer, a dual-branch Transformer-based architecture that jointly models temporal dynamics and system-level interactions to predict VM performance degradation. The temporal branch captures transient workload behavior at second-level granularity, while the system branch learns cross-metric dependencies. This design enables adaptation to both static and dynamic workloads without scenario-specific tuning, supported by a fine-grained dataset that significantly expands metric diversity and temporal detail compared to prior benchmarks.

In summary, this work makes the following contributions:
\begin{itemize}
    \item We introduce CloudFormer, a dual-branch transformer architecture that jointly models temporal and system-level dynamics for VM performance degradation prediction in black-box cloud environments.
    \item We integrate CloudPerfTrace\footnote{\label{foot:ds}This dataset is publicly available at \url{https://huggingface.co/datasets/AmirShahbaz/CloudPerfTrace}
    }, a rich dataset that captures 206 system metrics with a resolution of one second in diverse static and dynamic scenarios, allowing fine-grained modeling of transient interference effects.
    \item We demonstrate that CloudFormer consistently outperforms state-of-the-art baselines across multiple evaluation metrics, delivering robust generalization across diverse and previously unseen workloads. In particular, CloudFormer achieves a mean absolute error (MAE) of only 7.8\%, which represents a significant improvement in predictive accuracy over existing methods by at least 28\%.
    \item We provide an in-depth evaluation and ablation study that analyzes the contribution of each architectural branch, offering insights into the roles of temporal and system-level modeling. This not only validates design choices, but also demonstrates CloudFormer’s adaptability to varying workload patterns.
\end{itemize}

The remainder of this paper is structured as follows. Section~\ref{sec:related} reviews related work in the area of performance prediction. Section~\ref{sec:problem} formally defines the problem and introduces the key challenges addressed in this work. Section~\ref{sec:dataset} describes the dataset used in this study, including the collection methodology and the composition of the scenarios. Section~\ref{sec:cloudFormer} introduces CloudFormer, our proposed deep learning-based methodology to model performance degradation. Section~\ref{sec:exp} presents the experimental setup along with a detailed analysis of the results. Finally, Section~\ref{sec:conclusions} concludes the paper with a summary of the findings and potential directions for future work.

\section{Related Works}
\label{sec:related}

\subsection{VM Performance Challenges and Mitigation}

Virtual machines (VMs) are widely used in modern cloud environments to provide isolation and flexible resource management among co-located tenants. Traditionally, VM management strategies have focused on workload-based resource provisioning, relying on predicted load levels to optimize utilization and cost. However, managing VMs solely on the basis of workload does not account for performance degradation caused by resource interference among multiple VMs sharing the same physical host.
To address such degradation, prior works explored runtime performance-level management through heuristic prediction or sandbox-based cloning approaches~\cite{novakovic_deepdive_2013, vasic_dejavu_2012}, comparing the performance of a virtual machine with that of a clone running in a controlled environment. Although this approach can detect degradation, it incurs significant computational overhead and is impractical at scale. To reduce these costs, Wang et al.~\cite{wang_modeling_2016} proposed an analytical model to predict interference among co-running Apache Spark tasks, but it is limited to Spark and lacks generality. Similarly, ML-based runtime prediction methods, such as those in~\cite{shekhar_performance_2018}, focus on latency-sensitive Spark applications, restricting their applicability to broader cloud workloads.

Other approaches, including collaborative filtering~\cite{delimitrou_paragon_2013}, attempt to preemptively schedule applications to minimize interference. However, these methods operate only during deployment and cannot adapt to runtime dynamics, thus failing to address performance variations after initial placement. Source code-dependent approaches such as Aspen~\cite{spafford_aspen_2012}, Palm~\cite{tallent_palm_2014}, PEMOGEN~\cite{bhattacharyya_pemogen_2014}, and COMPASS~\cite{lee_compass_2015} achieve high accuracy but require internal application knowledge and runtime states. This reliance on detailed instrumentation makes them unsuitable for black-box public cloud scenarios. To overcome some of these limitations, Pham et al.~\cite{pham_predicting_2020} introduced a two-stage ML-based prediction method using runtime metadata; however, it only predicts execution time and does not address interference effects.

More recent methods have shifted toward explicit performance degradation analysis, but typically use classification rather than quantitative prediction. For example, ISOLATE~\cite{gu_identifying_2025} formulates performance monitoring as an anomaly detection task, identifying metric correlation violations to signal QoS issues. Horchulhack~\cite{horchulhack_detection_2024} classifies whether a VM is experiencing interference or operating normally. Although these methods can detect when degradation occurs, they do not forecast future performance metrics or quantify degradation levels, limiting their utility for proactive mitigation and fine-grained resource planning.

\subsection{Performance Prediction Methods}

Performance prediction in cloud computing has evolved significantly, employing diverse strategies ranging from statistical analyses and regression techniques to advanced neural network models. Despite considerable progress, existing methods exhibit certain limitations that impede their generalizability and practical applicability.

Table~\ref{table:method_comparison} provides a detailed comparison of prominent methods based on critical criteria, including A) the capability to handle interference between multiple applications, B) scenario variety (e.g., CPU-intensive, network-intensive, etc.), C) explicit prediction of performance degradation, D) modeling of temporal behavior, E) number of metrics utilized, F) applicability to black-box scenarios, and G) handling of unknown applications.

\begin{table}[h!]
\centering
\caption{Comparison of different methods across metrics A–G.}
\label{table:method_comparison}
\renewcommand{\arraystretch}{1.5}
\begin{tabular}{l|c|c|c|c|c|c|c}
\textbf{Method} & \textbf{A} & \textbf{B} & \textbf{C} & \textbf{D} & \textbf{E} & \textbf{F} & \textbf{G} \\
\hline
Cloud White~\cite{pons_cloud_2023} & \textcolor{red}{\xmark} & \textcolor{red}{\xmark} & \cmark & \textcolor{red}{\xmark} & \textcolor{red}{10} & \cmark & \textcolor{red}{\xmark} \\
\hline
Seq2Seq~\cite{buchaca_sequence--sequence_2020} & \textcolor{red}{\xmark} & \cmark & \cmark & \cmark & \textcolor{red}{9} & \cmark & \textcolor{red}{\xmark} \\
\hline
Rusty~\cite{masouros_rusty_2021} & \cmark & \cmark & \textcolor{red}{\xmark} & \cmark & \textcolor{red}{3} & \cmark & \cmark \\
\hline
Seer~\cite{gan_seer_2019} & \cmark & \cmark & \textcolor{red}{\xmark} & \cmark & \textcolor{red}{1} & \textcolor{red}{\xmark} & \textcolor{red}{\xmark} \\
\hline
Monitorless~\cite{grohmann_monitorless_2019} & \cmark & \cmark & \textcolor{red}{\xmark} & \textcolor{red}{\xmark} & 117 & \cmark & \cmark \\ \hline
CloudProphet~\cite{huang_cloudprophet_2024} & \cmark & \cmark & \cmark & \textcolor{red}{\xmark} & 53 & \cmark & \textcolor{red}{\xmark} \\
\hline
\textbf{Ours} & \cmark & \cmark & \cmark & \cmark & \textbf{206} & \cmark & \cmark \\
\end{tabular}

\end{table}

\textit{Cloud White}~\cite{pons_cloud_2023} leverages multivariate regression to predict latency degradation among latency-critical workloads. However, it is limited by its scenario specificity and the lack of explicit temporal modeling. The Seq2Seq model proposed by Buchaca et al.~\cite{buchaca_sequence--sequence_2020}, despite effectively capturing temporal behavior, focuses primarily on pairwise scenarios, restricting scalability to larger-scale multi-tenant environments. Although it does not explicitly output a single performance degradation metric, its proposed percentage completion (PC) feature can be adapted to serve as an indirect measure of slowdown or performance degradation in co-scheduled settings.

\textit{Rusty}~\cite{masouros_rusty_2021} employs long-short-term memory (LSTM) networks for fine-grained predictions but does not directly model QoS degradation explicitly. Similarly, \textit{Seer}~\cite{gan_seer_2019} relies on detailed application instrumentation, limiting its practicality in typical black-box scenarios from public clouds.
\textit{Monitorless}~\cite{grohmann_monitorless_2019} overcomes this limitation by relying solely on platform metrics to detect saturation; however, it is limited to binary classification, lacking the ability to quantify the magnitude of performance degradation.

\textit{CloudProphet}~\cite{huang_cloudprophet_2024} significantly advances performance prediction through neural networks and a degradation index but fails to model explicit temporal behaviors and still requires prior knowledge of the type of application.

Collectively, these methods exhibit shortcomings in addressing generalizability across diverse scenarios, dynamic workload handling, and the black-box nature typical in real-world cloud environments. Most approaches rely heavily on predefined scenarios (e.g., CPU, memory, network-intensive workloads), using only scenario-specific metrics. Consequently, they fail to generalize across mixed and dynamically evolving workload scenarios, hindering their practical deployment.

\subsection{Performance Monitoring and Prediction Datasets}
Several publicly available datasets have significantly contributed to performance monitoring and prediction research in cloud computing environments. Google’s Borg cluster traces~\cite{wilkes_more_2011, wilkes_google_2020}, Alibaba’s cluster dataset \cite{tian_characterizing_2019}, and Microsoft’s Azure Resource Central \cite{cortez_resource_2017} provide insights into resource usage at scale, covering thousands of nodes running various realistic workloads. However, these datasets typically offer limited granularity, ranging from minutes to hours, and focus exclusively on resource metrics such as CPU and memory usage, often lacking explicit performance indicators like throughput or latency. Moreover, workload specifics are usually abstracted, with no direct correlation between resource usage and application-level performance. The CloudProphet dataset \cite{huang_cloudprophet_2024} partially addresses these limitations by capturing application-level performance for five CloudSuite benchmarks~\cite{palit_demystifying_2016}, yet it remains restricted to approximately 250-hour experiments with fewer resource metrics.

In summary, existing works on VM performance management and prediction either focus on limited scenarios, rely on extensive instrumentation, or lack explicit temporal modeling and generalizability. Additionally, publicly available datasets often fall short in capturing fine-grained, diverse metrics necessary to train and evaluate robust models under realistic cloud dynamics. These methodological and data-related gaps highlight the urgent need for comprehensive, flexible, and generalizable approaches to the prediction of performance degradation. Addressing this need, our work introduces a novel dual-branch Transformer-based architecture capable of modeling both temporal and system-level interactions without relying on source code or scenario-specific tuning. Complemented by a rich, fine-grained dataset, our approach bridges critical gaps in existing research and provides a foundation for more reliable and adaptive cloud performance prediction.

\section{Problem Definition}
\label{sec:problem}

\begin{figure}[t]
    \centering
    \includegraphics[width=\linewidth]{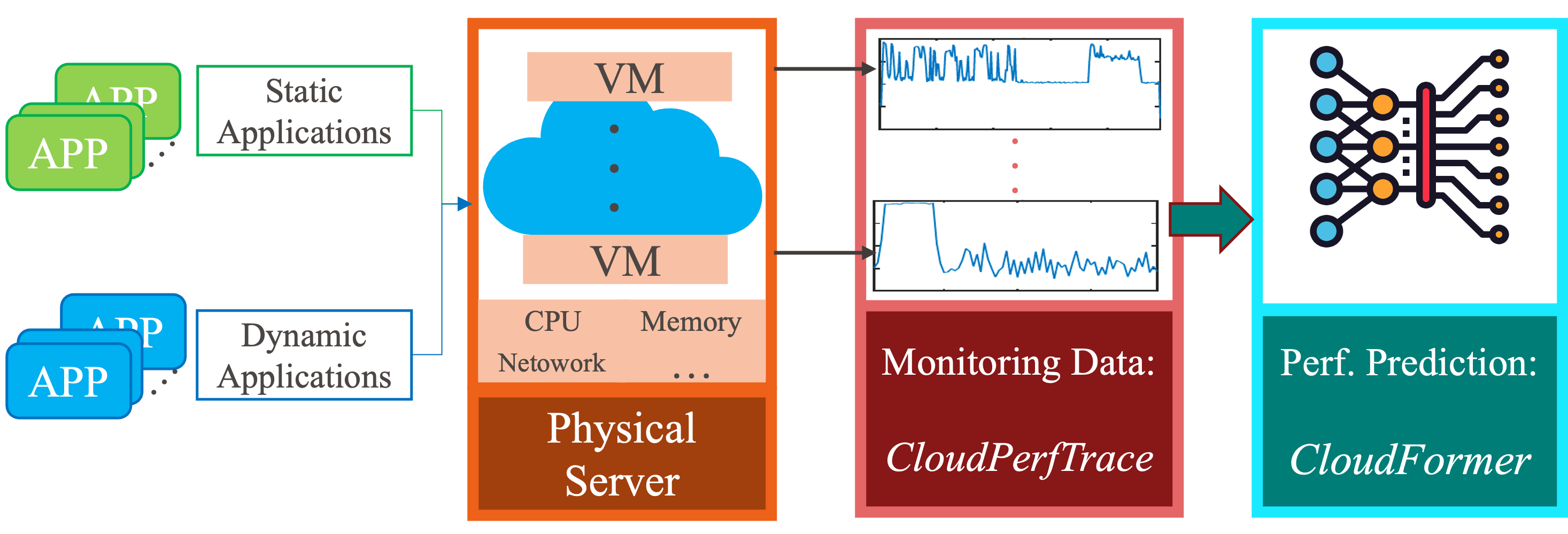}
    \caption{General overview of the performance degradation prediction problem in multi-tenant cloud environments. Multiple VMs compete for shared resources, causing interference and variable execution times. Due to the black-box nature of public clouds, only system-level traces observable from the host can be used for modeling and prediction.}
    \label{fig:problem_overview}
\end{figure}

Figure~\ref{fig:problem_overview} illustrates the general formulation of the performance prediction problem in multi-tenant cloud environments. In such environments, multiple virtual machines (VMs) are consolidated on the same physical server to maximize resource utilization. While virtualization technologies guarantee allocation of dedicated resources (e.g., vCPUs, memory quotas, storage partitions), they cannot ensure complete performance isolation. VMs inevitably compete for shared resources such as last-level cache (LLC), memory bandwidth, I/O subsystems, and network interfaces. This competition, referred to as performance interference, can lead to significant variability in execution times and throughput, even for identical applications running under nominally similar configurations.

The challenge is compounded by the black-box nature of public cloud environments. Cloud providers typically have no access to the source code or internal state of user applications due to privacy constraints. Therefore performance prediction must rely solely on system-level signals that are observable from outside the VM. These traces are inherently indirect and noisy, requiring robust modeling to accurately infer the performance impact of workload variations and resource contention.

In this study, our objective is to predict the performance of applications in cloud computing environments based on observed system-level metrics, where each application is executed alone within its own virtual machine. Each execution, whether the VM runs alone or in conjunction with other VMs, produces a data matrix \( x_i \in \mathbb{R}^{S \times T_i} \), where \( S \) is the number of system features and \( T_i \) is the number of time steps corresponding to its duration. Although all executions share the same number of features \( S \), the duration \( T_i \) may vary depending on the specific application and workload dynamics.
\hbl{Because performance degradation is non-linear and may occur at any point during execution, the model is designed to map the cumulative system-level interactions within} \( T_i \) \hbl{to a scalar performance index P.}
The performance \( \mathcal{P} \) for each execution is typically quantified as the ratio of the ideal Performance Metric \( PM_{\text{ideal}} \) to the actual observed Performance Metric \( PM_{\text{actual}} \). \hbl{To ensure a fair baseline for dynamic scenarios,} \( PM_{\text{ideal}} \) \hbl{is obtained by replaying the identical workload request-rate trace in an isolated environment (i.e., the VM running alone on the physical host), thus capturing the application's peak performance under that specific workload pattern without external interference.} Formally, this metric is defined as:
\begin{equation}
\mathcal{P} = \frac{PM_{\text{ideal}}}{PM_{\text{actual}}}
\end{equation}
where \( 0 < \mathcal{P} \leq 1 \). A value of \( \mathcal{P} = 1 \) corresponds to no degradation, while values approaching zero indicate significant performance degradation.
The definition of the performance metric may differ depending on the characteristics of the application or its execution context. These alternative formulations will be discussed in detail in Section~\ref{sec:dataset}.

Building on this problem formulation, we propose a novel scheme to predict performance degradation under these constraints. Our approach combines fine-grained system-level observations with advanced learning mechanisms to disentangle the effects of workload dynamics and interference. To support this, we introduce CloudPerfTrace (Section~\ref{sec:dataset}), a comprehensive dataset that captures system traces across diverse static and dynamic scenarios, and CloudFormer (Section~\ref{sec:cloudFormer}), a dual-branch Transformer architecture designed to jointly model temporal and system-level dependencies for accurate performance prediction.

\section{CloudPerfTrace\footref{foot:ds}: A Comprehensive Dataset for Performance Analysis}
\label{sec:dataset}

As discussed in Section~\ref{sec:related}, there is currently no comprehensive dataset that fully captures performance degradation in various scenarios. Our study presents a uniquely detailed dataset collected over two months from a test server, explicitly designed to explore diverse behaviors through systematically varied workloads. Unlike other datasets, we capture
system-level metrics at an unprecedented granularity of one-second intervals, enabling high-resolution analysis of transient system behaviors. Our dataset includes 206 different resource metrics gathered from Linux perf, hypervisor statistics, and Intel’s top-down analysis~\cite{yasin_top-down_2014}, significantly surpassing the feature richness of datasets such as CloudProphet (which covers fewer metrics in only five cloud applications). We also expand the workload coverage to 11 different cloud applications (which will be introduced later in Table \ref{tab:applications}), offering broader behavioral insights compared to previous work. This detailed, feature-rich, and temporally extensive dataset is particularly suited to fine-grained, machine learning-driven performance modeling, going beyond traditional utilization forecasting toward precise prediction of application-level performance dynamics.
Following the similar setup described in~\cite{huang_cloudprophet_2024}, we run multiple virtual machines concurrently to emulate realistic cloud environments. The following subsection details the experimental setup and application scenarios.

\subsection{Testbed Server Setup}
The experiments in this study are conducted on a test server equipped with two Intel Xeon Gold 6240 processors. Each processor has 36 virtual CPUs (vCPUs) and operates at a base frequency of 2.6 GHz. The server also features 384 GB of ECC memory and a 6 TB NVMe solid-state drive.

For the software setup, the server runs Ubuntu 20.04. Virtualization is managed using a combination of libvirt~\cite{libvirt_libvirt_nodate}, QEMU~\cite{qemu_qemu_nodate}, OpenvSwitch~\cite{openvswitch_open_nodate}, and KVM~\cite{kvm_kvm_nodate}. Each VM has allocated 4 vCPUs and 8 GB of memory, allowing each CPU socket to host up to 9 VMs concurrently. Therefore, any VM can experience random performance interference from the other 8 VMs located in the same CPU socket.

To simulate realistic scenarios in which server VMs receive requests from clients, the second CPU socket is dedicated to running 9 client VMs. The client and server VMs are interconnected via Open vSwitch. This setup ensures that server VMs are isolated from client-side interference while still allowing for performance interference among the server VMs themselves.

\subsection{Profiling Features}

\begin{table}[t]
\centering
\small
\caption{Representative list of typical collected hardware metrics and the number of features per category}
\label{table:libvirt_statistics}
    \begin{tabular}{ c c c } \toprule
    \textbf{Category} & \textbf{Count} & \textbf{Typical extracted metrics} \\
    \midrule
    \multirow{2}{*}{VM metrics} & \multirow{2}{*}{53} & CPU utilization level (\%) \\
      & & Unused Memory (KB)\\
      \midrule
    \multirow{3}{*}{Linux Perf} & \multirow{3}{*}{38} & Cycles (\#) \\
     & & LLC misses (\#) \\ 
     & & Retired instructions (\#) \\
      \midrule
   
    \multirow{2}{*}{Top-Down Analysis} & \multirow{2}{*}{12} & Frontend bound (\%) \\
     & & Backend bound (\%) \\
   \bottomrule
    \end{tabular}
\end{table}

\begin{table*}[ht]
\centering
\caption{Overview of application scenarios, purposes, performance metrics, total counts, percentage of static workload scenarios, and duration in days.}
\begin{tabular}{l c ccccccc}
\toprule
\textbf{Application}  & \textbf{Purpose}& \textbf{Performance Metric} & \textbf{Count} & \textbf{Static \%} & \textbf{Duration} & \textbf{Min. \%} & \textbf{Max. \%}\\
\midrule
Data Serving*    & Stress the data store and serving server & Operations/s     & 4209  & 65  & 16.43 & 67.16 & 100\\
Redis*           & Evaluate in-memory data structure server &Requests/s  & 4862  & 51  & 18.57 & 6.03 & 98.88 \\
Web Search*      & Simulate a search engine handling queries & Operations/s   & 2849  & 76  & 11.66  & 26.49 & 99.98\\
Graph Analytics & Analyze large-scale graph computations & Execution time   & 3195  & 100 & 49.69 & 85.76 & 98.37\\
Data Analytics  & Evaluate batch processing of large-scale datasets & Execution time  & 3085  & 100 & 118.95  &  87.63 & 97.47 \\
MLPerf          & Benchmark machine learning models & Requests/s & 3355  & 100 & 27.11 & 84.02 & 96.92 \\
Hbase           & Assess a distributed NoSQL database & Latency (s) & 4759  & 100 & 6.93  & 57.95 & 98.08\\
Alluxio*         & Evaluate memory-centric distributed storage system & Throughput & 3085  & 83  & 7.04 & 57.88 & 92.61\\
Minio*           & Evaluate lightweight object storage performance & Throughput  & 3139  & 82  & 7.09 & 10.65 & 99.46\\
TPCC            & Simulate a transaction processing systems & Latency (ms) & 2985  & 100 & 27.03 & 28.4 & 99.98 \\
Flink          & Benchmark real-time stream performance  & Operations/ms   & 3181  & 100 & 26.96 & 32.01 & 99.99\\
\bottomrule
\end{tabular}
\label{tab:applications}
\end{table*}

\begin{figure*}[h!]
    \centering
    \includegraphics[width=\linewidth]{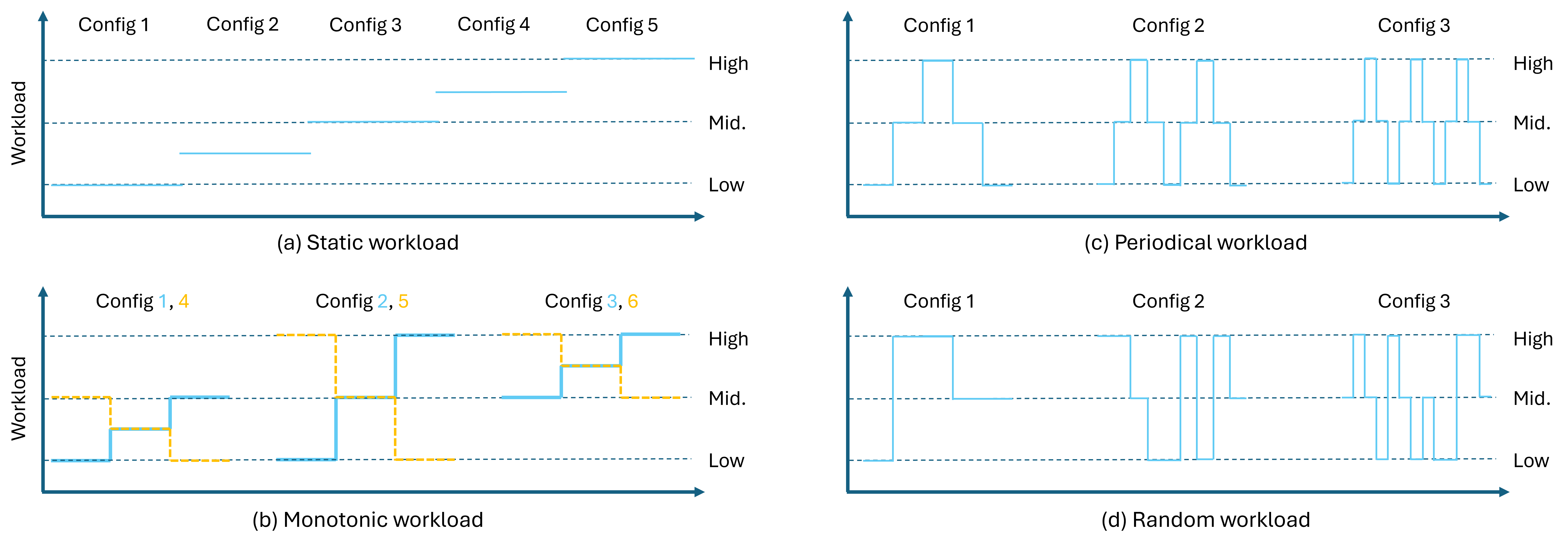} 
    \caption{For applications with dynamic workloads, four distinct scenarios are considered: (a) static workload, (b) monotonic workload, (c) periodic workload, and (d) random workload.}
    \label{fig:workload_configs}
\end{figure*}

In this work, we adopt the black-box scenario assumption, where cloud providers have no visibility into the applications running inside VMs due to privacy constraints. Consequently, we can only rely on basic hardware usage information that can be accessed from the host server (hypervisor) without breaching the VM’s internal state.

For example, 53 VM-level metrics such as CPU utilization and memory usage are collected using the libvirt API. 38 low-level hardware metrics, including retired instructions and last-level cache (LLC) misses, are gathered via the Linux perf tool. Additionally, we employ Intel's Top-Down Analysis Method to identify the additional 12 performance bottleneck metrics.

Utilizing these hardware metrics does not violate the black-box assumption as all the profiling is done from outside the VM, i.e., on the host server. Key representatives of these metrics are listed in Table~\ref{table:libvirt_statistics}.

The raw dataset provides performance metrics for each application at different timestamps on various virtual machines (VMs). This yields a total of 103 base features for the primary VM. Recognizing the importance of inter-VM interactions, we also compute the average of these 103 metrics across all concurrent neighborhood VMs. This aggregation doubles the feature space, resulting in a comprehensive set of 206 distinct system metrics (103 from the primary VM + 103 from the aggregated neighbors).

\subsection{Application Scenarios}

To comprehensively cover various cloud application scenarios, this work selects 11 representative cloud benchmarks, as listed in Table \ref{tab:applications}. In addition, cloud applications are inherently subject to varying workload levels, depending on incoming client requests. To capture this behavior, five of the applications (marked with an asterisk in Table \ref{tab:applications}) are configured with dynamic workload scenarios.

Figure \ref{fig:workload_configs} illustrates the workload configuration settings. As shown in Figure \ref{fig:workload_configs}(a), the static workload scenario represents the simplest case, where the application does not experience a variation in workload over time for different workload levels. Figures~\ref{fig:workload_configs}(b)–(d) depict the dynamic workload scenarios:
Figure \ref{fig:workload_configs}(b) presents the monotonic workload scenario, in which the workload either increases (blue line) or decreases (orange line) over time.
For more complex behavior, Figure \ref{fig:workload_configs}(c) depicts the periodic workload scenario, where the workload fluctuates in a regular repeating pattern in which three different frequencies of workload variation are considered. Finally, we include a random workload scenario to emulate unpredictable user request patterns that cloud applications may encounter in real-world settings (Figure \ref{fig:workload_configs}(d)).

\subsection{Data}

Table~\ref{tab:applications} summarizes the collected traces for all 11 applications, detailing the total number of runs, the proportion of static workload configurations, the cumulative recording duration, and the observed performance degradation range (Min. \% and Max. \%) for each application. Trace counts vary by application, reflecting differences in execution duration and scenario configurations. The percentages of static workload highlight that some applications (Graph Analytics, Data Analytics, MLPerf, HBase, TPCC, Flink) were exclusively evaluated under static conditions, while others (Data Serving, Redis, Web Search, Alluxio, Minio) were extensively tested with both static and dynamic scenarios (monotonic, periodic, and random). The duration column further indicates the total days of data recorded for each benchmark, with long-running experiments such as Data Analytics (118.95 days) and Graph Analytics (49.69 days) providing substantial coverage for steady-state and transient performance behaviors. The minimum and maximum performance percentages highlight the variability in degradation observed across runs, with some interactive applications like Redis and Minio experiencing severe degradation (down to $\sim$6-10\%), while other workloads like Data Analytics remained relatively stable.

In total, the dataset spans approximately 317 days of recorded traces across all applications, representing a comprehensive view of VM performance under diverse workloads. This diverse set of traces, which span thousands of runs, provides a rich dataset that captures both short-term variability and long-term workload trends. The breadth of applications, workload types, and temporal coverage make it well-suited for developing and evaluating generalizable performance prediction models in realistic multi-tenant cloud environments.

\section{CloudFormer: A Unified Approach to Black-Box VM Performance Prediction}
\label{sec:cloudFormer}

Herein, we introduce CloudFormer, a novel deep learning-based architecture designed to predict performance degradation in multi-tenant cloud environments. Performance degradation in such environments arises from two fundamentally different types of dynamics: (i) \textit{temporal dynamics}, describing how workloads evolve over time (e.g., sudden CPU bursts or periodic I/O spikes), and (ii) \textit{system-level interactions}, representing relationships among system metrics (e.g., Last-Level Cache occupation influencing the Cache Miss). These dynamics are highly nonlinear and often interdependent, making them challenging to capture with a single unified model.

\begin{figure}[t]
    \centering
    \includegraphics[width=\linewidth]{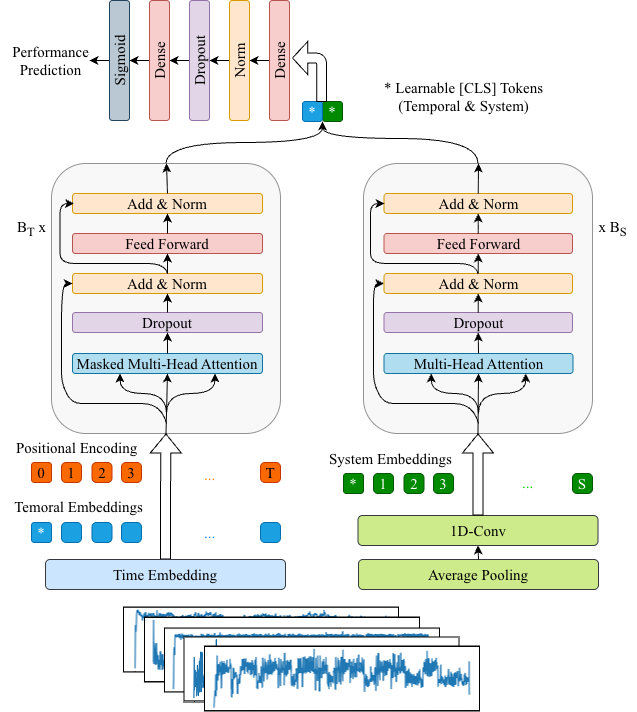}
    \caption{Overview of CloudFormer architecture illustrating dual branches for temporal and system-level modeling.}
    \label{fig:cf_overview}
\end{figure}

To address this, CloudFormer employs a dual-branch Transformer structure that explicitly separates temporal and system-level modeling. Figure~\ref{fig:cf_overview} illustrates the architecture. The input is processed in parallel through two branches: a \textit{temporal branch} (left), which models transient sequence dependencies using positional encoding and masked attention to accommodate variable-length executions; and a \textit{system branch} (right), which captures global cross-metric interactions by first aggregating the execution trace. This design explicitly separates the modeling of time-varying fluctuations from the holistic resource contention profile, allowing the system branch to efficiently learn metric dependencies without assuming any specific ordering.

This separation is intentional: in multi-tenant environments, evolving workloads and resource contention impact performance in distinct but complementary ways. By allowing each branch to specialize while maintaining a shared transformer-based encoder design, the model can capture rich domain-specific patterns. The outputs from both branches are merged through the prediction head, using complementary representations from temporal and system perspectives to produce accurate performance degradation forecasts. The following subsections provide a detailed description of its structure.

\subsection{Input Representation}

The input data is represented as a matrix of shape $S \times T$, where $S$ denotes the number of system metrics (e.g., CPU, memory, network, LLC miss), which is fixed across samples, and $T$ denotes execution time steps, which can vary per sample.

\subsection{Temporal Branch}

In the temporal branch, each time step is first normalized and mapped from the high-dimensional feature space (S=206) to a compact embedding space $d_t$ through a dense layer with ReLU activation. This learnable projection effectively distills the raw metrics into a concise latent representation, allowing the model to automatically extract salient features before temporal processing. A learnable class token is added to the sequence to capture a global temporal summary. To retain temporal order, sinusoidal positional encoding is added as follows:

\begin{equation}
\begin{aligned}
\text{PE}(pos, 2i) &= \sin\left(\frac{pos}{10000^{2i/d_t}}\right),\\
\text{PE}(pos, 2i+1) &= \cos\left(\frac{pos}{10000^{2i/d_t}}\right).
\end{aligned}
\end{equation}

This results in a shape embedding $(T+1) \times d_t$. The temporal branch then passes through $B_T$ transformer encoder blocks, each containing Masked Multi-Head Self-Attention (MHA) (to handle variable-length sequences with padding masks), dropout, residual connections with layer normalization, and position-wise feedforward layers. In this branch, the learnable class token effectively 'attends' to all time steps, aggregating relevant global temporal features into a single representation.

\subsection{System Branch}

For the system branch, the input is first averaged over the time dimension, effectively reducing temporal variability and summarizing each system feature. A 1D convolution then projects these pooled features into a system embedding space of dimension $d_s$. A learnable class token is prepended, resulting in an embedding of shape $(S+1) \times d_s$. No positional encoding is applied here, as system metrics are unordered by nature. The system branch then passes through $B_S$ Transformer encoder blocks, each comprising standard MHA (without masking), dropout, residual connections with layer normalization, and position-wise feed-forward layers. Here too, the learnable class token acts as a global summary vector, capturing interactions among system metrics.

\subsection{Attention Mechanism}

In both the temporal and system branches, the core operation for modeling dependencies is the self-attention mechanism~\cite{vaswani_attention_2017}, which enables each token to attend to every other token in the same branch. This mechanism computes a weighted sum of value vectors $V$, where the weights are determined by the similarity between query vectors $Q$ and key vectors $K$. Formally, the attention operation is defined as:
\begin{equation}
\text{Attention}(Q, K, V) = \text{softmax}\left(\frac{QK^T}{\sqrt{d}}\right)V,
\end{equation}
where $Q$, $K$, $V$ are learned linear projections of input embeddings, and $d$ is the dimensionality of each attention head.

The multi-head design further enhances representational power by learning multiple independent attention patterns in parallel, each focusing on different aspects of the data. The outputs are concatenated and passed through a feedforward projection, allowing the model to combine fine-grained relational cues with global contextual information. In both branches, the presence of a learnable class token ensures that the most salient information is aggregated into a single vector representation which will later be fused in the prediction head.

\subsection{Fusion and Prediction Head}

After the final encoder blocks, the representations corresponding to the class tokens of both branches are concatenated, forming a comprehensive joint embedding. This fused representation is passed through a multi-layer perceptron consisting of dense layers, normalization, dropout, and a final dense layer. In the end, a sigmoid activation is applied to produce a continuous scalar prediction of performance degradation, constrained between 0 and 1. The use of learnable class tokens ensures that each branch provides a highly expressive and global informed summary, enabling more accurate final predictions.

Overall, by combining temporal and system-level perspectives, CloudFormer is able to extract rich, complementary information, leading to superior prediction accuracy compared to single-branch models.

\section{Experiments}
\label{sec:exp}

\subsection{Experimental Setup}

We evaluated the model's generalization capabilities by partitioning the 11 applications (see Section~\ref{sec:dataset}) into seven training applications (four static, three dynamic) and four entirely unseen testing applications (two static, two dynamic). This balanced split ensures the model captures diverse system behaviors during training while being assessed under realistic, complex conditions. To mitigate selection bias, we repeated this randomized partitioning process six times and report the aggregate results.

\subsection{Implementation}

As discussed in Section~\ref{sec:cloudFormer}, CloudFormer explicitly separates temporal and system-level modeling. In the temporal branch, the input is projected in a dimension space $d_t = 64$, followed by the addition of a class token and positional encoding. A sequence of $B_T = 4$ transformer encoder blocks with masked Multi-Head Attention (MHA) handles variable-length time sequences. The system branch first applies mean pooling over time and projects them to a dimension $d_s = 64$ through a 1D convolution. This branch also prepends a class token but does not use positional encoding since the system metrics are unordered. The system features are then processed by $B_S = 4$ transformer encoder blocks with standard (unmasked) MHA. Finally, the class token outputs from both branches are concatenated and passed through an MLP head with Swish activation, layer normalization, and dropout (rate 0.4), producing a scalar output via a sigmoid activation. The Multi-Head Attention (MHA) modules were configured with a head size of 16 and four attention heads, while the feedforward network within each Transformer block was set to a dimension of 256, providing a good trade-off between model expressiveness and computational efficiency. Specifically, the resulting model is exceptionally lightweight, containing a total of $\sim228k$ trainable parameters. Of these, the temporal branch accounts for $\sim222k$ parameters, while the system branch is extremely compact with only $\sim6k$ parameters.

For training, the data was first normalized. The model was then trained using the Adam optimizer with a low initial learning rate of $1\mathrm{e}{-5}$ to ensure stable convergence. A logarithmic cosh loss function was employed to handle potential outliers smoothly. 
\hbl{The log-cosh loss was chosen over standard Mean Squared Error (MSE) because it is approximately equal to $\frac{x^2}{2}$ for small errors and $|x| - \log(2)$ for large errors, providing a robust and twice-differentiable alternative that is less sensitive to noise.}
To further improve generalization and efficiency, a learning rate scheduler combining linear warm-up and cosine decay was applied.

\subsection{Baseline Methods}

For benchmarking our proposed model, we considered five widely used regression methods: Decision Trees (DT)~\cite{cao_load_2018}, Random Forests (RF)~\cite{pham_predicting_2020, grohmann_monitorless_2019}, Linear Regression (LR), Gamma Regression with an Inverse Power Link function (GLR)~\cite{pons_cloud_2023}, and long-short-term memory networks (LSTM)~\cite{masouros_rusty_2021}. Decision Trees build prediction models by recursively partitioning the feature space to minimize prediction errors, resulting in straightforward predictive structures. Random forests extend this concept by creating an ensemble of decision trees trained on random subsets of features and data, enhancing prediction robustness and reducing variance.

Both Decision Trees and Random Forests underwent hyperparameter optimization using Bayesian search combined with cross-validation. Specifically, for Decision Trees, we optimized hyperparameters including maximum depth, minimum samples required at each leaf, minimum samples required for splitting, and feature selection criteria. For Random Forests, we optimized parameters such as the number of estimators (trees), minimum samples per leaf, and feature selection methods. Hyperparameter optimization ensures these baselines represent competitive and reliable benchmarks against which we can effectively compare our model's performance.

Additionally, we included Linear Regression as a simpler baseline method, which fits a linear relationship between the input features and the target variable. This method provides a fundamental performance reference point that assesses the predictive complexity of the dataset.

The Gamma Regression (GLR) baseline, implemented with an inverse power link function using a Generalized Linear Model, was also employed. GLR is particularly suitable for modeling strictly positive, continuous outcomes that exhibit skewness, making it a relevant baseline for comparison against more complex nonlinear models.

Lastly, we incorporated long-short-term memory networks (LSTM) to evaluate performance from a time-domain perspective. Unlike other baselines, LSTM models explicitly capture sequential dependencies, which makes them particularly effective for modeling temporal dynamics in the data. For evaluation, we thus considered both feature domain solutions (Decision Trees, Random Forests, Linear Regression, Gamma Regression) and a time-domain solution (LSTM), providing a comprehensive benchmarking framework for assessing our proposed model's effectiveness.

Finally, we note that while prominent works such as Seer~\cite{gan_seer_2019}, Seq2Seq~\cite{buchaca_sequence--sequence_2020}, and CloudProphet~\cite{huang_cloudprophet_2024} were discussed in Section~\ref{sec:related}, they are excluded from this quantitative benchmark due to fundamental differences in problem formulation. Seer and Monitorless formulate performance monitoring as a classification task (predicting the probability of QoS violations and resource saturation, respectively) rather than a regression of degradation. Similarly, Seq2Seq focuses on forecasting multivariate resource usage traces (e.g., generating future CPU and memory time-series) rather than predicting a unified scalar performance index. Lastly, CloudProphet relies on application-specific profiles from a known set of benchmarks, making it unsuitable for  our evaluation setting, where the model must generalize to entirely unknown applications.

\subsection{Results}

The performance results, summarized in Table~\ref{tab:mse_mae_comparison}, clearly demonstrate the effectiveness of the CloudFormer (CF) compared to baseline methods. CF achieved the lowest mean squared error (MSE) of $142.67 \pm 49.71$ and the lowest mean absolute error (MAE) of $7.80 \pm 1.55$, significantly outperforming all baseline methods by at least 28\%. Among the baseline models, Random Forests (RF) exhibited the next best performance, with an MSE of $205.67 \pm 94.82$ and MAE of $10.78 \pm 3.29$.

\begin{table}[t]
\centering
\caption{Performance comparison of methods (mean $\pm$ std)}
\begin{tabular}{lcc}
\toprule
\textbf{Method} & \textbf{MSE} & \textbf{MAE} \\\midrule
LR     & $5465.67 \pm 4047.98$       & $49.09 \pm 12.45$ \\
GLR~\cite{pons_cloud_2023}     & $1.05 \times 10^{10} \pm 2.58 \times 10^{10}$ & $477.12 \pm 1008.71$ \\
DT~\cite{cao_load_2018}& $419.67 \pm 183.68$         & $15.03 \pm 3.96$ \\
RF~\cite{pham_predicting_2020, grohmann_monitorless_2019}      & $205.67 \pm 94.82$          & $10.78 \pm 3.29$ \\
LSTM~\cite{masouros_rusty_2021}     & $427.67 \pm 228.93$          & $15.42 \pm 4.17$ \\
\midrule
CF  & $142.67 \pm 49.71$          & $7.80 \pm 1.55$ \\
\bottomrule
\end{tabular}
\label{tab:mse_mae_comparison}
\end{table}

In contrast, linear methods such as LR and GLR performed notably worse, underscoring the complexity and nonlinearity of the dataset. The time-domain method, LSTM, achieved moderate performance with an MSE of $427.67 \pm 228.93$ and an MAE of $15.42 \pm 4.17$, emphasizing the importance of capturing information in both the temporal and feature domains. In general, the superior performance of CF indicates that effective modeling of both temporal and system dynamics provides substantial benefits for accurate prediction of performance.

\begin{figure}[h!]
    \centering
    \includegraphics[width=0.9\linewidth]{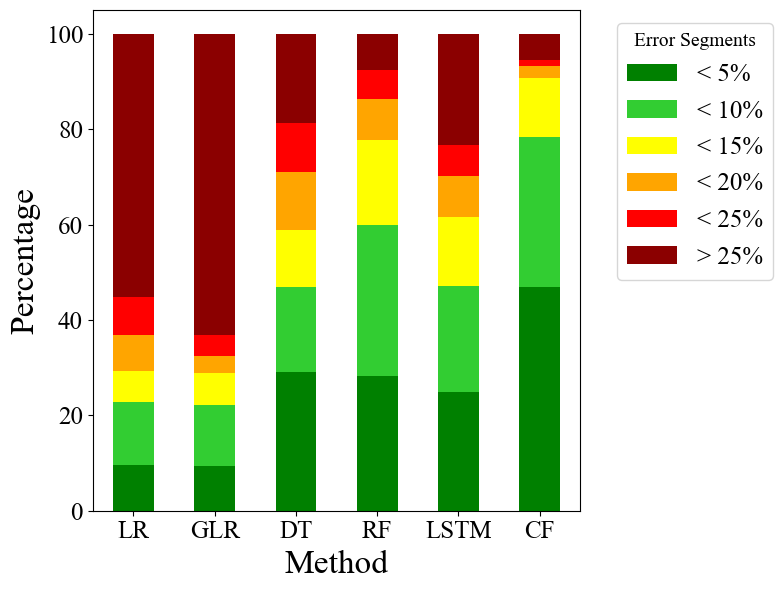}
    \caption{Stacked bar plot illustrating the distribution of prediction errors across different models, showing the percentage of samples falling within specific error segments.}
    \label{fig:performance_comparison}
\end{figure}

To provide a more interpretable view of the prediction error distributions, Figure~\ref{fig:performance_comparison} illustrates the percentage of predictions that fall within specific error bands (i.e., $<5\%$, $<10\%$, $<15\%$, $<20\%$, $<25\%$, and $>25\%$). This plot reveals a clear trend: CF exhibits the highest concentration of accurate predictions in the lower error bands, particularly below 5\% and 10\%, and the smallest proportion of large errors above 25\%. In contrast, linear models such as LR and GLR suffer from a large proportion of high-error predictions, while DT, RF, and LSTM show intermediate behavior. These results reinforce that CF's dual-domain attention design leads to both higher accuracy and more consistent prediction reliability across scenarios.

\subsection{Ablation Study}

\subsubsection{Impact of Temporal and System Branches}

To better understand the contribution of each domain-specific attention mechanism, we conducted an ablation study by independently evaluating the temporal and system branches of CloudFormer. Specifically, we constructed two variants of the model: CF-Temporal, including only the temporal self-attention branch, and CF-System, which includes only the system self-attention branch. The results of this study are presented in Table~\ref{tab:mse_mae_comparison_ablation}.

\begin{table}[t]
\centering
\caption{Comparing performance of individual branches and the full CF model (mean $\pm$ std)}
\begin{tabular}{lcc}
\toprule
\textbf{Method} & \textbf{MSE} & \textbf{MAE} \\\midrule
CF-Temporal  & $156.33 \pm 70.24$        & $8.47 \pm 1.89$ \\
CF-System  & $171.33 \pm 57.73$          & $8.65 \pm 1.55$ \\
CF  & $142.67 \pm 49.71$                 & $7.80 \pm 1.55$ \\
\bottomrule
\end{tabular}
\label{tab:mse_mae_comparison_ablation}
\end{table}

The results highlight the distinct and complementary roles of each branch. Notably, the CF-System branch achieves an MAE of $8.65 \pm 1.55$, outperforming the Random Forest baseline ($10.78 \pm 3.29$) despite containing only $\sim6k$ parameters. This demonstrates that the branch's dense embedding strategy effectively distills the 206 metrics into a concise global profile. While the CF-Temporal branch achieves a slightly lower MAE of $8.47 \pm 1.89$, it requires the majority of the model's capacity ($\sim222k$ parameters) to capture transient dynamics. Ultimately, the full CF model fuses these representations to achieve the best overall performance, confirming that integrating the efficient, global system-level context with fine-grained temporal dynamics enables the model to extract richer representations than either branch could achieve in isolation.

\subsubsection{Per-Application Error Analysis}

\begin{figure*}[h!]
    \centering
    \includegraphics[width=\linewidth]{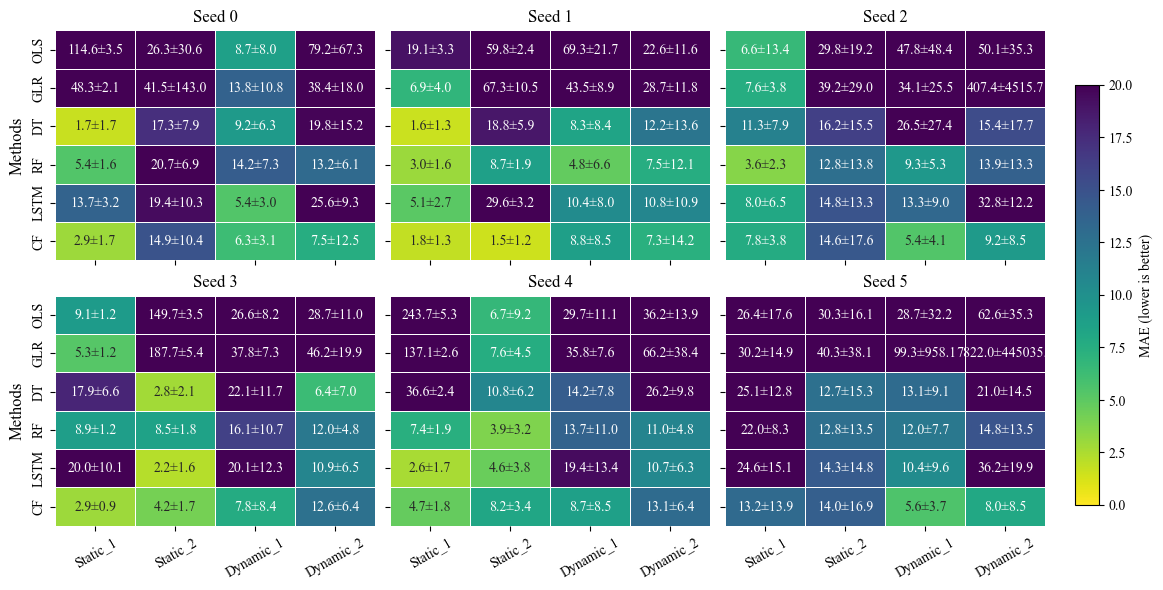} %
    \caption{Heatmaps showing the mean absolute error (MAE $\pm$ STD) of different models across six random seeds and four test applications (two static-only and two dynamic scenarios). For better visual contrast, the color bar is capped at 20\% MAE.}
    \label{fig:heatmap_results}
\end{figure*}

To better understand the robustness of the model and generalization at the application level, we conducted a detailed per-application error analysis on all seeds tested. Figure~\ref{fig:heatmap_results} illustrates the mean absolute error (MAE) performance of different models across six random seeds, evaluated on four applications per seed. In each heatmap, the rows correspond to the baseline methods and CloudFormer (CF), while the columns represent the four test applications (two static-only and two with dynamic scenarios). The color intensity encodes the MAE value for each method-application pair, where lighter colors indicate lower errors (better performance) and darker shades indicate higher errors. For better visual contrast, the color bar is capped at 20\% MAE.

As seen consistently in the last row of each subfigure, CloudFormer (CF) outperforms all other methods on average across all seeds and application scenarios. Although there are isolated instances where certain baseline methods achieve slightly lower MAE on specific applications, e.g., in Seed 2 on Static 1, where three methods show marginally better performance, these gains are not consistent across other applications and come at the cost of significantly higher errors elsewhere. This indicates that while some methods may overfit simpler static applications, they fail to generalize effectively to dynamic or more complex scenarios, resulting in poorer overall performance for that seed.

Furthermore, the heatmaps suggest that static scenarios are generally easier for models to predict, as evidenced by the lower error values compared to dynamic scenarios. This aligns with expectations: static applications involve more stable and predictable behavior, requiring the model to learn fewer complex relationships. However, dynamic scenarios introduce additional variability and resource interference, making prediction inherently more challenging. In particular, there are a few exceptions where certain static applications exhibit higher errors, for instance, in Seed 0 Static 2, Seed 2 Static 2, and Seeds 5 Static 1 and Static 2. These higher errors in static scenarios can be attributed to specific applications such as TPCC and Flink, which display more distinctive or irregular behavior compared to other static workloads. Overall, CF’s robust and consistent performance across both static and dynamic cases highlights its superior generalization capability and adaptability to complex cloud environments.

\section{Conclusions}
\label{sec:conclusions}

Ensuring reliable performance in multi-tenant cloud environments remains a critical challenge due to the dynamic nature of workloads and the complex resource contention among co-located virtual machines. Performance degradation not only undermines the quality of service guarantees, but also limits the efficiency of resource management strategies. Accurate prediction of degradation is therefore essential for proactive mitigation and optimized cloud operations.

In this work, we have introduced CloudFormer, a dual-branch Transformer-based architecture explicitly designed to address these challenges. CloudFormer separates temporal and system-level modeling to better capture the distinct sources of performance variability. The temporal branch focuses on workload evolution over time, while the system branch models cross-metric interactions that influence performance. These complementary representations are fused to provide a unified prediction of VM performance degradation.  

Our approach is supported by a rich dataset comprising 206 system metrics, collected at one-second resolution, across both static and dynamic workload scenarios. This dataset provides a high-fidelity view of both transient and long-term behaviors in realistic cloud environments.  
The experimental evaluation has shown that CloudFormer consistently outperforms state-of-the-art baselines across multiple metrics, achieving robust generalization across diverse workloads. In particular, the model has a mean absolute error of just 7.8\%, which highlights its predictive accuracy and reliability.

Looking ahead, future work could explore extending CloudFormer to multi-node settings, integrating online adaptation for changing workload patterns, and incorporating energy-efficiency objectives alongside performance prediction. These directions hold promise in enabling more intelligent, resource-aware cloud platforms capable of providing stronger performance guarantees at scale.

\section*{Acknowledgments}
\hbl{This work was supported in part by the HeatingBits S4S project of EPFL, Grants PID2021-126576NB-I00, PID2024-158311NB-I00 funded by MCIN/AEI/10.13039/501100011033 and FEDER, EU.}

\printbibliography

@inproceedings{vaswani_attention_2017,
	author = {Vaswani, Ashish and Shazeer, Noam and Parmar, Niki and Uszkoreit, Jakob and Jones, Llion and Gomez, Aidan N and Kaiser, L ukasz and Polosukhin, Illia},
	booktitle = {NeurIPS},
	publisher = {Curran Associates, Inc.},
	title = {Attention is {All} you {Need}},
	url = {https://proceedings.neurips.cc/paper_files/paper/2017/hash/3f5ee243547dee91fbd053c1c4a845aa-Abstract.html},
	urldate = {2025-08-13},
	year = {2017}}

@misc{kvm_kvm_nodate,
	author = {{KVM}},
	title = {https://linux-kvm.org/},
	url = {https://linux-kvm.org/},
	urldate = {2025-08-13}}

@misc{openvswitch_open_nodate,
	author = {{OpenVSwitch}},
	title = {https://www.openvswitch.org/},
	url = {https://www.openvswitch.org/},
	urldate = {2025-08-13}}

@misc{qemu_qemu_nodate,
	author = {{QEMU}},
	title = {https://www.qemu.org/},
	url = {https://www.qemu.org/},
	urldate = {2025-08-13}}

@misc{libvirt_libvirt_nodate,
	author = {{Libvirt}},
	title = {libvirt: {The} virtualization {API}},
	url = {https://libvirt.org/},
	urldate = {2025-08-13}}

@misc{gartner_gartner_2024,
	author = {Gartner},
	journal = {Gartner},
	language = {en},
	month = nov,
	title = {Gartner {Forecasts} {Worldwide} {Public} {Cloud} {End}-{User} {Spending} to {Total} \$723 {Billion} in 2025},
	url = {https://tinyurl.com/bdfzkz8d},
	urldate = {2025-07-29},
	year = {2024}}

@misc{iea_ai_2025,
	author = {IEA},
	journal = {IEA},
	language = {en-GB},
	month = apr,
	title = {{AI} is set to drive surging electricity demand from data centres while offering the potential to transform how the energy sector works},
	url = {https://tinyurl.com/c6pnaft3},
	urldate = {2025-07-29},
	year = {2025}}

@inproceedings{cao_load_2018,
	author = {Cao, Rui and Yu, Zhaoyang and Marbach, Trent and Li, Jing and Wang, Gang and Liu, Xiaoguang},
	booktitle = {{COMPSAC}},
	doi = {10.1109/COMPSAC.2018.00109},
	month = jul,
	note = {ISSN: 0730-3157},
	pages = {728--737},
	title = {Load {Prediction} for {Data} {Centers} {Based} on {Database} {Service}},
	url = {https://ieeexplore.ieee.org/document/8377734},
	urldate = {2025-08-06},
	volume = {01},
	year = {2018}}

@inproceedings{palit_demystifying_2016,
	author = {Palit, Tapti and Shen, Yongming and Ferdman, Michael},
	booktitle = {{ISPASS}},
	doi = {10.1109/ISPASS.2016.7482080},
	month = apr,
	pages = {122--132},
	title = {Demystifying cloud benchmarking},
	url = {https://ieeexplore.ieee.org/document/7482080},
	urldate = {2025-08-06},
	year = {2016}}

@inproceedings{yasin_top-down_2014,
	author = {Yasin, Ahmad},
	booktitle = {{ISPASS}},
	doi = {10.1109/ISPASS.2014.6844459},
	month = mar,
	pages = {35--44},
	title = {A {Top}-{Down} method for performance analysis and counters architecture},
	url = {https://ieeexplore.ieee.org/abstract/document/6844459},
	urldate = {2025-08-05},
	year = {2014}}

@article{wood_sandpiper_2009,
	author = {Wood, Timothy and Shenoy, Prashant and Venkataramani, Arun and Yousif, Mazin},
	doi = {10.1016/j.comnet.2009.04.014},
	issn = {1389-1286},
	journal = {Computer Networks},
	month = dec,
	pages = {2923--2938},
	series = {Virtualized {Data} {Centers}},
	shorttitle = {Sandpiper},
	title = {Sandpiper: {Black}-box and gray-box resource management for virtual machines},
	url = {https://www.sciencedirect.com/science/article/pii/S1389128609002035},
	urldate = {2025-08-04},
	year = {2009}}

@misc{amd_amd_nodate,
	author = {{AMD} {Solutions} for {HCI} and {Virtualization}},
	journal = {AMD},
	language = {en},
	title = {https://www.amd.com/en/solutions/hci-and-virtualization.html},
	url = {https://www.amd.com/en/solutions/hci-and-virtualization.html},
	urldate = {2025-07-30}}

@article{neiger_intel_2006,
	author = {Neiger, G and Santoni, A and Leung, F and Rodgers, D and Uhlig, R},
	doi = {10.1535/itj.1003},
	issn = {1535-864X},
	journal = {Intel Technology Journal},
	month = jun,
	number = {03},
	title = {Intel{\textregistered} {Virtualization} {Technology}},
	volume = {10},
	year = {2006}}

@article{akbar_game-based_2021,
	author = {Akbar, Saeed and Malik, Saif Ur Rehman and Choo, Kim-Kwang Raymond and Khan, Samee U. and Ahmad, Naveed and Anjum, Adeel},
	doi = {10.1109/TCC.2019.2899310},
	issn = {2168-7161},
	journal = {IEEE Transactions on Cloud Computing},
	month = jul,
	number = {3},
	pages = {845--853},
	title = {A {Game}-based {Thermal}-{Aware} {Resource} {Allocation} {Strategy} for {Data} {Centers}},
	url = {https://ieeexplore.ieee.org/document/8642435},
	urldate = {2025-07-29},
	volume = {9},
	year = {2021}}

@article{anwar_game-theoretic_2021,
	author = {Anwar, Ahmed H. and Atia, George and Guirguis, Mina},
	doi = {10.1109/TCC.2019.2905605},
	issn = {2168-7161},
	journal = {IEEE TCC},
	month = jul,
	number = {3},
	pages = {854--867},
	title = {A {Game}-{Theoretic} {Framework} for the {Virtual} {Machines} {Migration} {Timing} {Problem}},
	url = {https://ieeexplore.ieee.org/document/8668476},
	urldate = {2025-07-29},
	volume = {9},
	year = {2021}}

@article{kim_virtual_2013,
	author = {Kim, Shin-Gyu and Eom, Hyeonsang and Yeom, Heon Y.},
	doi = {10.1007/s11227-013-0939-2},
	issn = {0920-8542},
	journal = {J. Supercomput.},
	month = dec,
	number = {3},
	pages = {1489--1506},
	title = {Virtual machine consolidation based on interference modeling},
	url = {https://doi.org/10.1007/s11227-013-0939-2},
	urldate = {2025-07-29},
	volume = {66},
	year = {2013}}

@misc{mytton_data_2022,
	author = {Mytton, David},
	journal = {Dev Sustainability},
	language = {en},
	month = feb,
	title = {Data center energy and {AI} in 2025. In: Dev Sustainability.},
	url = {https://www.devsustainability.com/p/data-center-energy-and-ai-in-2025},
	urldate = {2025-07-29},
	year = {2022}}

@inproceedings{delimitrou_paragon_2013,
	author = {Delimitrou, Christina and Kozyrakis, Christos},
	doi = {10.1145/2451116.2451125},
	isbn = {978-1-4503-1870-9},
	month = mar,
	pages = {77--88},
	publisher = {Association for Computing Machinery},
	series = {{ASPLOS} '13},
	shorttitle = {Paragon},
	title = {Paragon: {QoS}-aware scheduling for heterogeneous datacenters},
	url = {https://dl.acm.org/doi/10.1145/2451116.2451125},
	urldate = {2025-07-08},
	year = {2013}}

@inproceedings{shekhar_performance_2018,
	author = {Shekhar, Shashank and Abdel-Aziz, Hamzah and Bhattacharjee, Anirban and Gokhale, Aniruddha and Koutsoukos, Xenofon},
	booktitle = {{IEEE} {CLOUD}},
	doi = {10.1109/CLOUD.2018.00018},
	month = jul,
	note = {ISSN: 2159-6190},
	pages = {82--89},
	title = {Performance {Interference}-{Aware} {Vertical} {Elasticity} for {Cloud}-{Hosted} {Latency}-{Sensitive} {Applications}},
	url = {https://ieeexplore.ieee.org/document/8457786},
	urldate = {2025-07-08},
	year = {2018}}

@inproceedings{lee_compass_2015,
	author = {Lee, Seyong and Meredith, Jeremy S. and Vetter, Jeffrey S.},
	doi = {10.1145/2751205.2751220},
	isbn = {978-1-4503-3559-1},
	month = jun,
	pages = {405--414},
	publisher = {Association for Computing Machinery},
	series = {{ICS} '15},
	shorttitle = {{COMPASS}},
	title = {{COMPASS}: {A} {Framework} for {Automated} {Performance} {Modeling} and {Prediction}},
	url = {https://dl.acm.org/doi/10.1145/2751205.2751220},
	urldate = {2025-07-08},
	year = {2015}}

@inproceedings{bhattacharyya_pemogen_2014,
	author = {Bhattacharyya, Arnamoy and Hoefler, Torsten},
	booktitle = {{PACT}},
	doi = {10.1145/2628071.2628100},
	month = aug,
	pages = {393--404},
	shorttitle = {{PEMOGEN}},
	title = {{PEMOGEN}: {Automatic} adaptive performance modeling during program runtime},
	url = {https://ieeexplore.ieee.org/document/7855916},
	urldate = {2025-07-08},
	year = {2014}}

@inproceedings{tallent_palm_2014,
	author = {Tallent, Nathan R. and Hoisie, Adolfy},
	doi = {10.1145/2597652.2597683},
	isbn = {978-1-4503-2642-1},
	month = jun,
	pages = {221--230},
	publisher = {Association for Computing Machinery},
	series = {{ICS} '14},
	shorttitle = {Palm},
	title = {Palm: easing the burden of analytical performance modeling},
	url = {https://dl.acm.org/doi/10.1145/2597652.2597683},
	urldate = {2025-07-08},
	year = {2014}}

@inproceedings{spafford_aspen_2012,
	author = {Spafford, Kyle L. and Vetter, Jeffrey S.},
	booktitle = {{SC} '12},
	doi = {10.1109/SC.2012.20},
	month = nov,
	note = {ISSN: 2167-4337},
	pages = {1--11},
	shorttitle = {Aspen},
	title = {Aspen: {A} domain specific language for performance modeling},
	url = {https://ieeexplore.ieee.org/document/6468530},
	urldate = {2025-07-08},
	year = {2012}}

@inproceedings{wang_modeling_2016,
	author = {Wang, Kewen and Khan, Mohammad Maifi Hasan and Nguyen, Nhan and Gokhale, Swapna},
	booktitle = {{IEEE} {CLOUD}},
	doi = {10.1109/CLOUD.2016.0063},
	note = {ISSN: 2159-6190},
	pages = {423--431},
	title = {Modeling {Interference} for {Apache} {Spark} {Jobs}},
	url = {https://ieeexplore.ieee.org/document/7820300},
	urldate = {2025-07-08},
	year = {2016}}

@article{horchulhack_detection_2024,
	author = {Horchulhack, Pedro and Viegas, Eduardo K. and Santin, Altair O. and Ramos, Felipe V. and Tedeschi, Pietro},
	copyright = {https://www.elsevier.com/tdm/userlicense/1.0/},
	doi = {10.1016/j.jnca.2024.103839},
	issn = {1084-8045},
	journal = {JNCA},
	language = {en},
	month = apr,
	note = {Publisher: Elsevier BV},
	pages = {103839},
	title = {Detection of quality of service degradation on multi-tenant containerized services},
	url = {https://linkinghub.elsevier.com/retrieve/pii/S108480452400016X},
	urldate = {2025-07-07},
	volume = {224},
	year = {2024}}

@article{gu_identifying_2025,
	author = {Gu, Wenwei and Liu, Jinyang and Chen, Zhuangbin and Zhang, Jianping and Su, Yuxin and Gu, Jiazhen and Feng, Cong and Yang, Zengyin and Yang, Yongqiang and Lyu, Michael R.},
	doi = {10.1145/3702978},
	issn = {1049-331X},
	journal = {ACM TOSEM},
	month = feb,
	number = {3},
	pages = {64:1--64:31},
	title = {Identifying {Performance} {Issues} in {Cloud} {Service} {Systems} {Based} on {Relational}-{Temporal} {Features}},
	url = {https://dl.acm.org/doi/10.1145/3702978},
	urldate = {2025-07-07},
	volume = {34},
	year = {2025}}

@inproceedings{tian_characterizing_2019,
	address = {New York, NY, USA},
	author = {Tian, Huangshi and Zheng, Yunchuan and Wang, Wei},
	doi = {10.1145/3357223.3362710},
	isbn = {978-1-4503-6973-2},
	month = nov,
	pages = {139--151},
	publisher = {Association for Computing Machinery},
	series = {{SoCC} '19},
	title = {Characterizing and {Synthesizing} {Task} {Dependencies} of {Data}-{Parallel} {Jobs} in {Alibaba} {Cloud}},
	url = {https://dl.acm.org/doi/10.1145/3357223.3362710},
	urldate = {2025-04-23},
	year = {2019}}

@inproceedings{cortez_resource_2017,
	address = {Shanghai China},
	author = {Cortez, Eli and Bonde, Anand and Muzio, Alexandre and Russinovich, Mark and Fontoura, Marcus and Bianchini, Ricardo},
	booktitle = {26th {Symposium} on {Operating} {Systems} {Principles}},
	doi = {10.1145/3132747.3132772},
	isbn = {978-1-4503-5085-3},
	language = {en},
	month = oct,
	pages = {153--167},
	publisher = {ACM},
	shorttitle = {Resource {Central}},
	title = {Resource {Central}: {Understanding} and {Predicting} {Workloads} for {Improved} {Resource} {Management} in {Large} {Cloud} {Platforms}},
	url = {https://dl.acm.org/doi/10.1145/3132747.3132772},
	urldate = {2025-04-23},
	year = {2017}}

@misc{wilkes_more_2011,
	author = {Wilkes, John},
	month = nov,
	title = {More {Google} cluster data},
	type = {Google research blog},
	url = {https://research.google/blog/more-google-cluster-data/},
	year = {2011}}

@techreport{wilkes_google_2020,
	address = {Mountain View, CA, USA},
	author = {Wilkes, J},
	institution = {Google Inc.},
	title = {Google cluster-usage traces v3},
	type = {Technical {Report}},
	url = {https://github.com/google/cluster-data},
	year = {2020}}

@article{huang_cloudprophet_2024,
	author = {Huang, Darong and Costero, Luis and Pahlevan, Ali and Zapater, Marina and Atienza, David},
	doi = {10.1109/TSUSC.2024.3359325},
	issn = {2377-3782},
	journal = {IEEE TSC},
	month = jul,
	number = {4},
	pages = {661--676},
	shorttitle = {{CloudProphet}},
	title = {{CloudProphet}: {A} {Machine} {Learning}-{Based} {Performance} {Prediction} for {Public} {Clouds}},
	url = {https://ieeexplore.ieee.org/abstract/document/10415550},
	urldate = {2025-04-09},
	volume = {9},
	year = {2024}}

@inproceedings{gan_seer_2019,
	author = {Gan, Yu and Zhang, Yanqi and Hu, Kelvin and Cheng, Dailun and He, Yuan and Pancholi, Meghna and Delimitrou, Christina},
	doi = {10.1145/3297858.3304004},
	isbn = {978-1-4503-6240-5},
	month = apr,
	pages = {19--33},
	publisher = {Association for Computing Machinery},
	series = {{ASPLOS} '19},
	shorttitle = {Seer},
	title = {Seer: {Leveraging} {Big} {Data} to {Navigate} the {Complexity} of {Performance} {Debugging} in {Cloud} {Microservices}},
	url = {https://doi.org/10.1145/3297858.3304004},
	urldate = {2024-07-15},
	year = {2019}}

@article{buchaca_sequence--sequence_2020,
	author = {Buchaca, David and Marcual, Joan and Berral, Josep LLuis and Carrera, David},
	doi = {10.1016/j.future.2020.03.058},
	issn = {0167-739X},
	journal = {FGCS},
	month = sep,
	pages = {155--166},
	title = {Sequence-to-sequence models for workload interference prediction on batch processing datacenters},
	url = {https://www.sciencedirect.com/science/article/pii/S0167739X19310921},
	urldate = {2024-07-15},
	volume = {110},
	year = {2020}}

@article{masouros_rusty_2021,
	author = {Masouros, Dimosthenis and Xydis, Sotirios and Soudris, Dimitrios},
	doi = {10.1109/TPDS.2020.3013948},
	issn = {1558-2183},
	journal = {IEEE TPDS},
	month = jan,
	note = {Conference Name: IEEE Transactions on Parallel and Distributed Systems},
	number = {1},
	pages = {184--198},
	shorttitle = {Rusty},
	title = {Rusty: {Runtime} {Interference}-{Aware} {Predictive} {Monitoring} for {Modern} {Multi}-{Tenant} {Systems}},
	url = {https://ieeexplore.ieee.org/document/9158547},
	urldate = {2024-07-15},
	volume = {32},
	year = {2021}}

@article{pons_cloud_2023,
	author = {Pons, Luc{\'\i}a and Feliu, Josu{\'e} and Sahuquillo, Julio and G{\'o}mez, Mar{\'\i}a E. and Petit, Salvador and Pons, Julio and Huang, Chaoyi},
	doi = {10.1016/j.future.2022.08.012},
	issn = {0167-739X},
	journal = {FGCS},
	month = jan,
	pages = {13--25},
	shorttitle = {Cloud {White}},
	title = {Cloud {White}: {Detecting} and {Estimating} {QoS} {Degradation} of {Latency}-{Critical} {Workloads} in the {Public} {Cloud}},
	url = {https://www.sciencedirect.com/science/article/pii/S0167739X22002734},
	urldate = {2024-06-24},
	volume = {138},
	year = {2023}}

@article{pham_predicting_2020,
	author = {Pham, Thanh-Phuong and Durillo, Juan J. and Fahringer, Thomas},
	doi = {10.1109/TCC.2017.2732344},
	issn = {2168-7161},
	journal = {IEEE TCC},
	month = jan,
	note = {Conference Name: IEEE Transactions on Cloud Computing},
	number = {1},
	pages = {256--268},
	title = {Predicting {Workflow} {Task} {Execution} {Time} in the {Cloud} {Using} {A} {Two}-{Stage} {Machine} {Learning} {Approach}},
	volume = {8},
	year = {2020}}

@article{vasic_dejavu_2012,
	author = {Vasi{\'c}, Nedeljko and Novakovi{\'c}, Dejan and Miu{\v c}in, Svetozar and Kosti{\'c}, Dejan and Bianchini, Ricardo},
	doi = {10.1145/2189750.2151021},
	issn = {0163-5964},
	journal = {ACM SIGARCH Computer Architecture News},
	month = mar,
	number = {1},
	pages = {423--436},
	shorttitle = {{DejaVu}},
	title = {{DejaVu}: accelerating resource allocation in virtualized environments},
	url = {https://dl.acm.org/doi/10.1145/2189750.2151021},
	urldate = {2023-07-28},
	volume = {40},
	year = {2012}}

@inproceedings{novakovic_deepdive_2013,
	author = {Novakovi{\'c}, Dejan and Vasi{\'c}, Nedeljko and Novakovi{\'c}, Stanko and Kosti{\'c}, Dejan and Bianchini, Ricardo},
	booktitle = {USENIX},
	isbn = {978-1-931971-01-0},
	language = {en},
	pages = {219--230},
	shorttitle = {\{{DeepDive}\}},
	title = {{DeepDive}: {Transparently} {Identifying} and {Managing} {Performance} {Interference} in {Virtualized} {Environments}},
	url = {https://www.usenix.org/conference/atc13/technical-sessions/presentation/novakovi%C4%87},
	urldate = {2023-07-28},
	year = {2013}}

@inproceedings{grohmann_monitorless_2019,
	address = {New York, NY, USA},
	author = {Grohmann, Johannes and Nicholson, Patrick K. and Iglesias, Jesus Omana and Kounev, Samuel and Lugones, Diego},
	booktitle = {20th {Int.} {Middleware} {Conference}},
	doi = {10.1145/3361525.3361543},
	isbn = {978-1-4503-7009-7},
	month = dec,
	pages = {149--162},
	publisher = {Association for Computing Machinery},
	series = {Middleware '19},
	shorttitle = {Monitorless},
	title = {Monitorless: {Predicting} {Performance} {Degradation} in {Cloud} {Applications} with {Machine} {Learning}},
	url = {https://dl.acm.org/doi/10.1145/3361525.3361543},
	urldate = {2023-07-17},
	year = {2019}}

\newpage
\section*{Artifact Appendix for ``CloudFormer: An Attention-based Performance Prediction for Public Clouds with Unknown Workload''}

\subsection{abstract}
This artifact provides the \textit{CloudPerfTrace} dataset, a collection of 206 system-level metrics captured at a one-second resolution over 317 days. Specifically designed for black-box multi-tenant cloud environments where internal VM telemetry is unavailable. The dataset enables precise prediction of performance degradation. It encompasses 11 diverse application tasks and captures the complex interplay between intrinsic workload variations and external resource interference.

\subsection{Artifact Summary}
\begin{itemize}
    \item \textbf{Dataset Title:} CloudPerfTrace: A High-Resolution Dataset for VM Performance Prediction.
    \item \textbf{Persistent Identifier:} \url{https://doi.org/10.57967/hf/7847}
    \item \textbf{Repository:} \url{https://huggingface.co/datasets/AmirShahbaz/CloudPerfTrace}
    \item \textbf{License:} Creative Commons Attribution 4.0 International (CC-BY-4.0).
    \item \textbf{Persistence Statement:} This artifact is permanently archived with a unique DOI to ensure guaranteed persistence and open access for research reproducibility.
\end{itemize}

\subsection{Schema and Features}
The dataset models system-level interactions by providing 206 metrics, split equally between the target VM and its concurrent neighbors. All features are collected from the host-level hypervisor to respect the black-box privacy constraints of public clouds.
\begin{itemize}
    \item \textbf{perf\_ori}: The target performance ratio ($0 < \mathcal{P} \le 1$) representing observed vs. ideal performance.
    \item \textbf{tr\_self / tr\_oth}: 53 VM-level metrics (e.g., CPU/Memory utilization) via libvirt.
    \item \textbf{lin\_self / lin\_oth}: 38 hardware counters (e.g., LLC misses, cycles) via Linux perf.
    \item \textbf{td\_self / td\_oth}: 12 Intel Top-Down analysis metrics.
\end{itemize}

\subsection{Dataset Context and Format}
To efficiently handle the large dataset collected over 317 days, we utilized the Apache Parquet format with Hive-style partitioning, which resulted in a compressed storage size of 18~GB.
\begin{itemize}
    \item \textbf{Partitioning Strategy:} Data is organized by \texttt{tasks} (Application ID), enabling the filtering and loading of specific workloads without full-memory residency.
    \item \textbf{Temporal Resolution:} 1-second intervals, providing the granularity needed to analyze transient interference effects.
    \item \textbf{Coverage:} 317 days of recording across 11 applications, including static, monotonic, periodic, and random workloads.
\end{itemize}

\subsection{Accessibility and Usage}
The mapping below allows users to utilize predicate pushdown to filter specific benchmarks evaluated in the paper.

\setcounter{table}{5}
\begin{table}[htbp]
\caption{Task ID to Application Mapping}
\begin{center}
\begin{tabular}{|c|l||c|l|}
\hline
\textbf{ID} & \textbf{Application} & \textbf{ID} & \textbf{Application} \\ \hline
4 & Data Serving & 11 & HBase \\ \hline
5 & Redis & 13 & Alluxio \\ \hline
6 & Web Search & 14 & Minio \\ \hline
7 & Graph Analytics & 15 & TPC-C \\ \hline
9 & Data Analytics & 16 & Flink \\ \hline
10 & MLPerf & & \\ \hline
\end{tabular}
\end{center}
\end{table}

\subsubsection{Environment Setup}
Accessing the dataset requires \texttt{pyarrow} and \texttt{huggingface\_hub} for high-performance columnar processing:
\begin{verbatim}
pip install pyarrow huggingface_hub
\end{verbatim}

\subsubsection{Downloading and Loading the Dataset}
The use of Hive-partitioning via PyArrow is recommended for performance:
\begin{verbatim}
import pyarrow.dataset as ds
from huggingface_hub import snapshot_download
repo_path = snapshot_download(
    repo_id="AmirShahbaz/CloudPerfTrace", 
    repo_type="dataset"
)

dataset = ds.dataset(
    f"{repo_path}/parquet_ds", 
    format="parquet", 
    partitioning="hive"
)
\end{verbatim}

Data can also be loaded partially, such as the Web Search data (Task ID 6):
\begin{verbatim}
#Ensure pandas is installed to use this command
web_search_df = dataset.to_table(
    filter=ds.field("tasks") == 6
).to_pandas()
\end{verbatim}

\end{document}